# EFFICIENT DISTRIBUTION GRID FLEXIBILITY PROVISION THROUGH MODEL-BASED MV GRID AND MODEL-LESS LV GRID APPROACH

*Ankur Majumdar[1,*], Omid Alizadeh-Mousavi[1]*

[1]*Depsys, Puidoux, Switzerland*
*ankur.majumdar@depsys.com

## Abstract

In order to maintain security and quality of supply while supporting increased intermittent generation and electrification of heating and transport at the LV grid, the provision of flexibility in the framework of distribution grid operation is pivotal. However, the availability of updated and accurate LV grid model is a challenge for the distribution system operator (DSO). This paper demonstrates a methodology of efficient flexibility provision and activation through a sensitivity coefficients-based model-less LV grid and a modelled MV grid approach. The paper further illustrates the value of monitoring and control through a LV DERMS platform for efficient realisation of this flexibility provision and activation. The methodology has been tested on a real MV and LV network of a Swiss DSO. The results show that there is a reduction of cost of operation for a DSO.

## 1  Introduction

With the increasing number of intermittent distributed energy resources (DERs) being connected at the medium (MV) and low voltage (LV) distribution grid, the distribution grid operator (DSO) has an increasing challenge of maintaining security and quality of supply. This means there is an increasing need to solve voltage and congestion issues through activation of flexibilities at the MV and LV levels. These flexibilities are available from a large number of resources such as, small-scale renewable generations, electric vehicles, energy storage, interruptible loads [1].

However, a large portion of these resources are not monitored by the DSO and as a result, a large portion of flexibilities available from the LV grid remains untapped. Furthermore, the flow of electricity from unmonitored and uncontrolled resources may create security issues in the distribution grid [2]. This requires improved capability to monitor MV and LV grids and DERs connected at the LV grid and control of the distributed resources to achieve optimal operation of the grid.

The DSOs are going through a process of smart meter rollout [3],[4]. The smart meter rollout program was initially considered to be promising for the grid monitoring and operation purposes however, it has many challenges. Firstly, the DSOs will have to bear a high cost of enabling fast data transfer and high-resolution data storage [5]. This means that the smart meters can only communicate the data once or twice a day to the central meter data management system. The smart meters measurements can be used by the DSO for billing purposes and notifications are used largely for fault localisation. Secondly, the data privacy issues associated with smart meter data renders the data to be used only after anonymising or pseudonymising them. Therefore, it is difficult in terms of practicality for the smart meter data to be used for flexibility or real-time operational decision purposes and hence, there is a need to have a grid monitoring equipment which can communicate control setpoints to the distributed and controlled resources.

Although, the DSO possesses all the grid topology information in its geographic information system (GIS), the access to an up-to-date and trustable information on LV grid topology and parameters to achieve the optimal operation of the grid for the DSO is difficult [6]. Therefore, recently, there has been focus on model-less approach for LV grid analysis.

This paper proposes a DSO flexibility operation platform with DER management system (DERMS) based on grid monitoring and control equipment at the secondary of the MV/LV transformers and the LV street cabinets and DERs. The DERMS can communicate control setpoints from the central DSO operation and management platform to the LV DERs. The implementation of the flexibility platform entails – i) aggregate *flexibility estimation* available from the LV grid through a model-less sensitivity coefficient approach. This ensures secure operation of the LV grid without requiring up to date topology and parameters of the LV grid; ii) *control and activation* of flexibilities of LV DERs from the central DSO operation platform through a DERMS platform. This is developed by a model-based MV OPF to ensure a secure DSO operation.

## 2. Proposed methodology for grid operation

*2.1 Current practice*

Today's distribution remains unmonitored at the LV level. At the MV level, the HV/MV substation has monitoring only at the feeder-outs. Thus, the MV grid remains largely unmonitored as well [7]. Due to the lack of monitoring information, the realised flexibility from the distribution grid is limited. The current philosophy of distribution grid operation is based on load allocation proportional to the kVA ratings of the MV/LV transformers [8]. However, with more and more intermittent generations being connected at the distribution grid, the security (voltage and line flows) of distribution grid is at stake. This requires improved visibility and control of the distributed resources. As the distribution grid operation is transitioning from passive to active, the voltage and line flows must be managed in an optimal way.

The optimal grid operation is a non-linear analysis based on the grid topology. However, the lack of access to the up-to-date and accurate topology information on the LV grid makes the model-based analysis extremely difficult. Therefore, LV model-less analysis based on sensitivity coefficients is advocated in some literatures to address this issue [9]. On the other hand, the up-to-date model of the MV grid is easily available in the GIS/NIS.

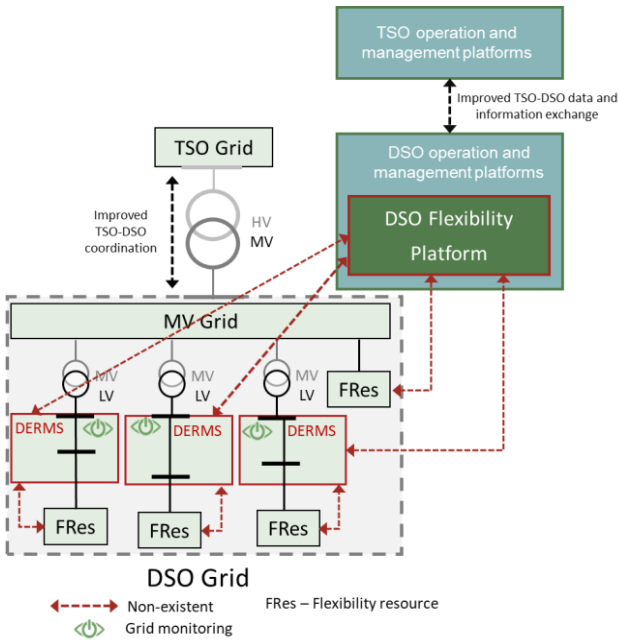

Fig. 1. The DSO operation with LV DERMS

*2.2. Proposed approach – model-less LV and model-based MV grid*

The proposed method consists of developing an operation and management platform, as shown in the Fig. 1, for the DSO. This central platform can communicate control setpoints to the LV DERMS at the MV/LV transformers and subsequently, to the DERs. The LV DERMS based on a model-less sensitivity coefficient approach estimates the available aggregate flexibilities at the MV/LV transformers from the LV grid. This methodology is facilitated by real-time and accurate monitoring data from grid monitoring equipment at the secondary of the MV/LV transformers.

The proposed framework for MV and LV grids with LV DERMS is realised by an optimal power flow (OPF) with real-time, accurate information of grid status which requires the MV grid model only. The control problem for DSO operation has two stages – a) MV level control, based on an ***MV OPF***, requires an MV grid model; b) The control of DERs at LV grid, achieved through ***LV OPF*** (LV DERMS) based on sensitivity coefficients, is incorporated in the ***MV OPF*** as constraints thus, removing the need for an up-to-date LV grid model.

The ***LV OPF*** is carried out first to estimate the aggregate flexibilities available from the LV grid. Once the flexibilities are estimated, the ***MV OPF*** is then carried out for the MV grid with estimated flexibility curves for all the LV grids as constraints.

Model-less ***LV OPF*** for aggregate flexibility estimation

The estimation of flexibility area from the flexibilities available at the LV side is a linear optimal power flow problem with model-less sensitivity coefficients. The formulation is described below.

$$\max \quad \alpha \Delta p_{sl}^{LV} + \beta \Delta q_{sl}^{LV} \quad (1)$$
$$\text{s. to} \quad \Delta V_i^{LV} = K_{VP} * \Delta P_i^{LV} + K_{VQ} * \Delta Q_i^{LV} \quad (2)$$
$$\Delta I_{ij}^{LV} = K_{IP} * \Delta P_i^{LV} + K_{IQ} * \Delta Q_i^{LV} \quad (3)$$
$$V_i^{LV} = V_{i,0}^{LV} + \Delta V_i^{LV} \quad (4)$$
$$I_{ij}^{LV} = I_{ij,0}^{LV} + \Delta I_{ij}^{LV} \quad (5)$$
$$V^{min} \leq V_i^{LV} \leq V^{max} \quad (6)$$
$$\left|I_{ij}\right|^{LV} \leq I_{max} \quad (7)$$

Eq (1) is the objective function maximising either active or reactive power flow through the MV/LV transformer. Eq. (2)-(7) are the linearised power flow constraints based on sensitivity coefficients $K_{VP}$, $K_{VQ}$, $K_{IP}$, $K_{IQ}$, (partial derivatives) calculated around the operating point with corresponding voltage $V_{i,0}^{LV}$ and branch current $I_{ij,0}^{LV}$, with respect to changes in active and reactive power injection or withdrawal at the LV nodes monitored with grid monitoring equipment. The flexibility curve is calculated by running the linear LV OPF several times depending on the number of points on the P-Q flexibility curve. The values of α and β can be $0, \pm 1$ depending on the search direction of P-Q flexibility curve through the MV/LV transformer [10],[11].

Model-based ***MV OPF*** for activation of flexibility

The methodology is a multi-objective framework for the DSOs with *three* objectives to address – **1)** Reduce their costs for security violations – minimise the costs for voltage violations and line flow violations (congestion); **2)** Reduce the technical operational losses – minimise the costs for losses in kWh; **3)** Reduce the costs for penalties on deviated schedules between the TSO and DSO – minimise the deviation of schedules agreed a priori with the TSO.

The problem formulation for the ***MV OPF*** is described below.

$$\min w_I \sum_{lines} r_{ij} l_{ij} + w_V \sum_{nodes} V_{dev} + w_{Ilim} \sum_{lines} l_{ij_{dev}} + w_p p_{sl}^{MV} + w_q q_{sl}^{MV} \quad (8)$$

s.to
$$\sum_{adj} P_{ij} + p_i = p_i^g - p_i^c \; \forall \, node, \forall \, adj \, lines \quad (9)$$
$$\sum_{adj} Q_{ij} + q_i = q_i^g - q_i^c \; \forall \, node, \forall \, adj \, lines \quad (10)$$
$$v_j = v_i - 2r_{ij}P_{ij} - 2x_{ij}Q_{ij} + (r_{ij}^2 + x_{ij}^2)l_{ij} \quad (11)$$
$$P_{ij}^2 + Q_{ij}^2 \leq v_i l_{ij} \quad (12)$$
$$p_i, q_i \in A_i^{MV/LV} \quad (13)$$
$$V_{dev} = 0 \; if \; V^{min} \leq V_i \leq V^{max} \quad (14)$$
$$V_{dev} = |V_i|^2 - V_{lim}^2, if \; V_i < V^{min} \; or \; V_i > V^{max} \quad (15)$$
$$\left|I_{ij}\right|_{dev}^2 = 0 \; if \left|I_{ij}\right| \leq I_{max} \quad (16)$$
$$\left|I_{ij}\right|_{dev}^2 = \left|I_{ij}\right|^2 - \left|I_{ij}\right|_{max}^2 \; if \left|I_{ij}\right| > I_{max} \quad (17)$$

Eq (8) is the objective function consisting of losses and the penalty costs for voltage and flow violations and the costs for active and reactive power flow violations at the slack bus (HV/MV transformer), which is the interface transformer between the TSO and the DSO. Eq. (9)-(12) represent the power flow constraints for the radial MV grid with linear DistFlow, with $v_i = |V_i|^2$ and $l_{ij} = |I_{ij}|^2$, and convexified second-order conic formulation [12]. Eq (13) constrains the control variables from the MV grid OPF within the linearised flexibility area from the model-less **LV OPF**. Eq(14)-(17) are the piece-wise linear penalty costs for voltage and flow violations.

## 3    Results

The proposed approach has been validated in three use cases given below.

- **Base case** – There is no control or monitoring of LV DERs
- **Monitoring case** – There are monitoring information on LV DERs but there is no control available.
- **Control case** – There is monitoring and control available through LV DERMS

The **Base case** represents the DSO operation with no grid monitoring information available. The setpoints are calculated based on the MV/LV transformers been allocated loads according to the rating of the transformers. The loads are allocated with the same profile as that of the MV feeder they are connected in. A power flow is then performed with the OPF setpoints to calculate the losses and the costs for security violations. In the **Monitoring case**, the DSO operation is carried out with grid monitoring information at the secondary of the MV/LV transformers. This case is simulated with an OPF for the MV grid. The grid monitoring devices are not capable to communicate control setpoints to the individual LV DERs connected downstream. While the **Control case** represents the DSO operation with the DERMS based platform for the LV grids, where the grid monitoring devices at the MV/LV transformers are capable to host DERMS platform. This case is simulated by sensitivity coefficients-based LV OPF for flexibility estimation for each LV grid; and the second-order conic OPF for the MV grid such that each setpoint lies within the estimated flexibility area of each MV/LV transformer, as explained in the previous section.

Table 1. Objectives and control variables for the **MV OPF** in each case

|  | Base case | Monitoring case | Control case |
|---|---|---|---|
| Objectives | Losses, violations | Losses, violations | Losses, violations |
| Control variables | MV slack voltage | MV slack voltage | MV slack voltage, Δp, Δq for controlled buses |

For clarity, the objectives and the control variables for **MV OPF** for the three cases are shown in Table 1.

Fig. 2 shows the test MV grid of a Swiss DSO with several grid monitoring devices located at the secondary of the MV/LV transformers. Fig. 3 represents an LV grid of such a network below a certain MV/LV transformer with installed grid monitoring devices. The grid monitoring devices provide highly accurate 10-min measurement data of voltage, current, active and reactive powers.

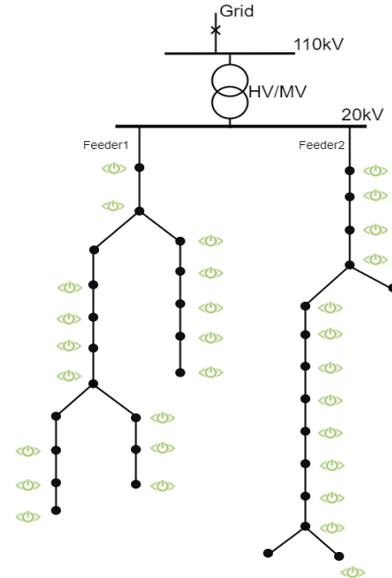

Fig. 2. A MV grid of a Swiss DSO

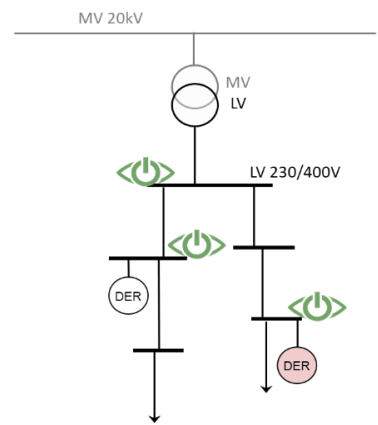

Fig. 3 A representative LV grid of the Swiss DSO

Fig. 4 further illustrates the consumption profiles of each MV feeder of the test network. This feeder profile is particularly useful in the **Base case** where there is no monitoring information available at the secondary of the MV/LV transformers and at the LV grids.

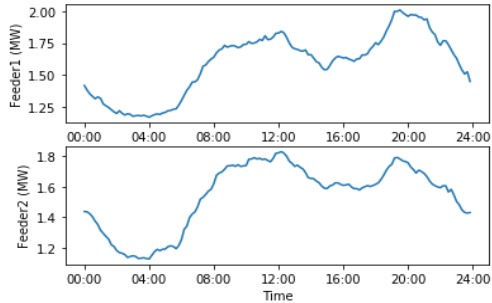

Fig. 4. Consumption profiles for each MV feeder

The test network has a total 200kWp of solar PVs connected to the grid at the LV level. The methodology is illustrated for a *current loading scenario* and a scenario with additional loads due to electrification of heating, cooling and transport – known as *future loading scenario*.

Table 2 and Fig 5 show the technical losses in kWh for 24 hours and the costs for security violation (calculated in CHF with the violations from their limits being multiplied by 100 CHF) for Base case, Monitoring case and Control case in the *current loading scenario*. It is found that there is a significant decrease in losses for Control case compared to that in Monitoring (5.26%) and Base case (5.27%). There are no security violation costs in the *current loading scenario* for the three cases. Moreover, the Control case proposed here includes a linear sensitivity coefficients-based *LV OPF* to estimate the aggregate flexibilities (based on 10% PV curtailment with pf control) from the LV grids at the MV/LV transformers as shown in Fig 6. The Monitoring and Base case have no information on the aggregate flexibility availability at the MV/LV transformers. Moreover, it can also be inferred that the hosting capacity (amount of new production that can be connected to the grid) of DERs increases in the Control case. Due to control through LV DERMS, the Control case provides a measure (200 kWp) for hosting capacity. However, there is no measure as to how much new production or consumption can be connected for the Monitoring and Base case due to no control of LV DERs or no LV monitoring information, respectively.

Table 2 KPIs for the three use cases in the current loading scenario

| KPIs | Base Case | Monitoring Case | Control Case |
|---|---|---|---|
| Grid violation costs (CHF) | 0 | 0 | 0 |
| Losses (kWh) | 159.43 | 159.42 | 151.03 |
| Flex at MV/LV (kW) | 0 | 0 | 20 |
| Hosting capacity of DERs (kWp) | - | - | 200 |

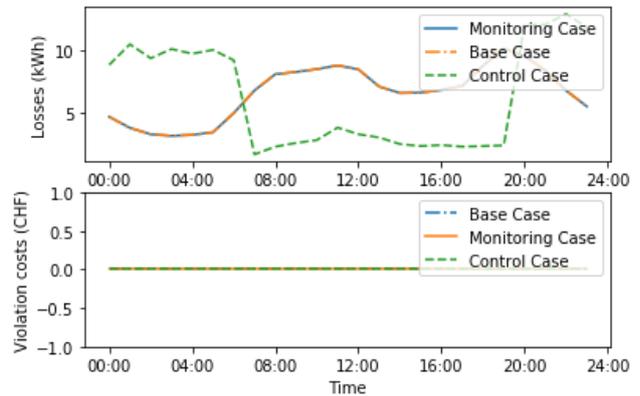

Fig. 5. Technical losses and security violation costs for three cases in the current loading scenario

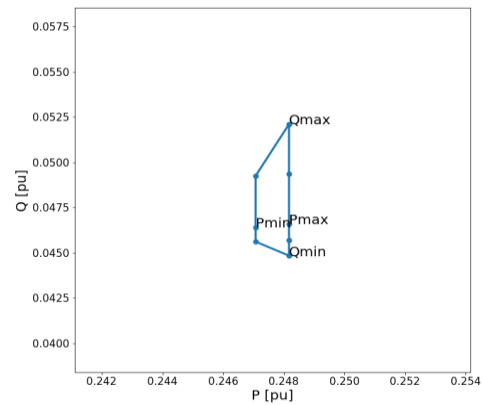

Fig. 6 P-Q flexibility capability estimation at MV/LV transformers

Fig. 7 and Table 3 show the comparison of total losses in kWh and the security violation costs for the three test cases in the *future loading scenario*, where there are additional loads of 500kW at each LV grid. The results further show that in the Control case there is a significant decrease in both losses (31%, 61%) and security violation costs (67%, 99.9%) compared to the Monitoring and Base case, respectively.

Fig. 8 further illustrates the zoomed in comparison of losses and security violation costs for Control and Monitoring cases in the *future loading scenario*.

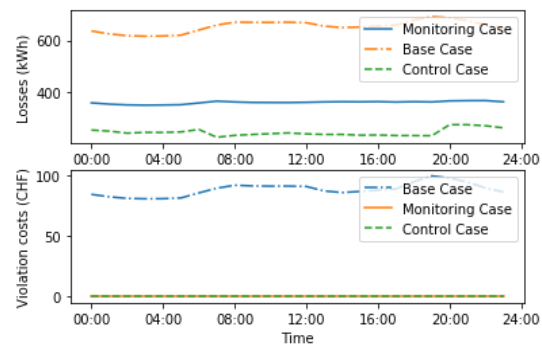

Fig. 7. Technical losses and security violation costs for three cases in the future loading scenario

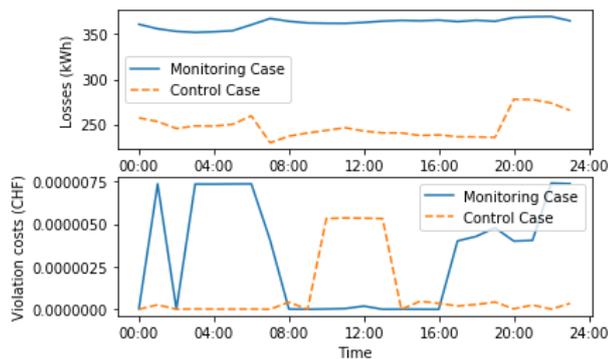

Fig. 8. Technical losses and security violation costs for Monitoring and Control Cases

Fig. 6 and Fig, 8 show that when the PV produces power the flow in the MV grid reduces thus, reducing the losses.

Table 3. KPIs for three cases for future loading scenario

| KPIs | Base Case | Monitoring Case | Control Case |
|---|---|---|---|
| Grid violation costs (CHF) | 2131.21 | 0.00077 | 0.00025 |
| Losses (kWh) | 15617.23 | 8700.22 | 5957.14 |
| Flex at MV/LV (kW) | 0 | 0 | 20 |
| Hosting capacity of DERs (kWp) | - | - | 200 |

The results infer that it is necessary to have LV DERMS based platform, capable to communicate control setpoints to the individual DERs, to realise the value of flexibility and for optimal operation of the MV grid.

## 4  Conclusion

The idea of optimal grid operation with DERMS at the MV/LV transformers is particularly beneficial for the DSO. The DSOs will be able to reduce costs while maintaining security and quality of supply. It will improve the hosting capacity of renewables in the DSO grid. Besides, the DSO will not require to have an up-to-date knowledge of their LV grid for their operation. This will further improve the flexibility potential of a DSO from the LV grid. The test case setup and the simulation results illustrate that LV DERMS can significantly improve the costs of operation and the security of the distribution grid. This will provide a DSO roadmap for the access of flexibilities to enable the transition from a passive to a decarbonised, decentralised and digitalised system.

## 5  Acknowledgements

This project is supported by the European Union's Horizon 2020 programme under the Marie Sklodowska-Curie grant agreement no. 840461.